\documentclass{ws-procs975x65}
\newcommand{\be}{\begin{equation}}
\newcommand{\ee}{\end{equation}}
\newcommand{\bea}{\begin{eqnarray}}
\newcommand{\eea}{\end{eqnarray}}
\usepackage[spanish,english]{babel}
\usepackage[latin1]{inputenc}

\newcommand{\R}{\mathcal{R}}
\newcommand{\Q}{\mathcal{Q}}

\begin{document}

\title{{\bf Enriched Phenomenology in Extended Palatini Theories}}

\author{Gonzalo J. Olmo}

\address{Instituto de Estructura de la Materia, CSIC, Serrano 121, 28006 Madrid, Spain}

\author{Hèlios Sanchis-Alepuz}
\address{ Fachbereich Theoretische Physik, Institut für Physik, Karl-Franzens-Universität Graz,
Universitätsplatz 5, A-8010 Graz, Austria}

\author{Swapnil Tripathi}
\address{Physics Department, University of Wisconsin-Barron County, 1800 College Drive, Rice Lake, Wisconsin 54868, USA}

\begin{abstract}
We show that extended theories of gravity with Lagrangian $f(R,R_{\mu\nu}R^{\mu\nu})$ in the Palatini formulation
possess a phenomenology much richer than the simpler $f(R)$ or $f(R_{\mu\nu}R^{\mu\nu})$ theories. In fact,
we find that the scalars $R$ and $Q=R_{\mu\nu}R^{\mu\nu}$ can be written as algebraic functions of the energy density and pressure of the energy momentum tensor. In the simpler cases of $f(R)$ or $f(R_{\mu\nu}R^{\mu\nu})$ theories, $R$ and $Q$ are just functions of the trace $T$ of the energy-momentum tensor. As a result, in radiation dominated universes $f(R)$ and $f(R_{\mu\nu}R^{\mu\nu})$ theories exhibit the same dynamics as general relativity with an effective cosmological constant. This is not the case of $f(R,R_{\mu\nu}R^{\mu\nu})$ models, in which $R=R(\rho,P)$ and $Q=Q(\rho,P)$ and, therefore, modified dynamics exists even for traceless sources.   
\end{abstract}

\bodymatter
\vspace{1cm}

In the recent literature on modified theories of gravity, $f(R)$ theories have received considerable attention. These theories are usually studied in two different forms, namely, in metric and in Palatini formalisms (see \cite{Capozziello:2009nq,Sotiriou:2008rp} and references therein). The former assumes that metric and connection are compatible, while the latter assumes that the connection is independent of the metric, i.e., it is not the Levi-Cività connection of the metric. Though extended theories of the form $f(R,Q)$, with $Q=R_{\mu\nu}R^{\mu\nu}$, have been thoroughly studied in the literature in metric formalism because of their relation with perturbative approaches to quantum gravity (semiclassical approach) and with low-energy limits of string theories, a similar analysis of the Palatini version of such theories is still lacking. This analysis is now motivated by the recent connections found between Palatini theories and non-perturbative approaches to quantum gravity\cite{Olmo-Singh}. The phenomenological scrutiny of these Palatini theories has also been limited due to the existence of certain technical difficulties related to the connection field equation. Those technical aspects were detailed and solved in a previous work\cite{OSAT09} and will be summarized here, where we discuss the properties of a fully tractable model and discuss the richness of its phenomenology.\\
The action that defines a Palatini $f(R,R_{\mu\nu}R^{\mu\nu})$ theory is as follows
\begin{equation}\label{eq:action}
S[g,\Gamma,\psi_m]=\frac{1}{2\kappa^2}\int d^4x \sqrt{-g}f(R,Q) +S_m[g,\psi_m]
\end{equation}
where $g_{\alpha\beta}$ represents the space-time metric, $\Gamma^\alpha_{\beta\gamma}$ is the connection (which is independent of the metric),  $\psi_m$ represents the matter fields, $R=g^{\mu\nu}R_{\mu\nu}$, and $Q=R_{\mu\nu}R^{\mu\nu}$. The Ricci tensor $R_{\mu\nu}={R_{\mu\rho\nu}}^\rho$ is defined in terms of the connection as $
R_{\mu\nu}(\Gamma )=-\partial_{\mu}
\Gamma^{\lambda}_{\lambda\nu}+\partial_{\lambda}
\Gamma^{\lambda}_{\mu\nu}+\Gamma^{\lambda}_{\mu\nu}\Gamma^{\rho}_{\rho\lambda}-\Gamma^{\lambda}_{\nu\rho}\Gamma^{\rho}_{\mu\lambda}$. Variation with respect to metric and connection lead to the following equations
\begin{eqnarray}
f_R R_{\mu\nu}-\frac{f}{2}g_{\mu\nu}+2f_QR_{\mu\alpha}R^\alpha_\nu &=& \kappa^2 T_{\mu\nu}\label{eq:met-varX}\\
\nabla_{\beta}\left[\sqrt{-g}\left(f_R g^{\mu\nu}+2f_Q R^{\mu\nu}\right)\right]&=&0
 \label{eq:con-varX}
\end{eqnarray}
were we have used the short-hand notation $f_R\equiv \partial_R f$, and $f_Q\equiv \partial_Q f$.  
The connection can be solved assuming the existence of an auxiliary metric $h_{\mu\nu}$ such that (\ref{eq:con-varX})
takes the form $\nabla_\mu\left[\sqrt{-h}h_{\alpha\beta}\right]=0$. Taking the energy-momentum tensor as a perfect fluid, we find 
\begin{equation}
h_{\mu\nu}=\Omega\left( g_{\mu\nu}-\frac{\Lambda_2}{\Lambda_1-\Lambda_2} u_\mu u_\nu \right) \ , \
h^{\mu\nu}=\frac{1}{\Omega}\left( g^{\mu\nu}+\frac{\Lambda_2}{\Lambda_1} u^\mu u^\nu \right)
\end{equation}
where $\Omega=\left[\Lambda_1(\Lambda_1-\Lambda_2)\right]^{1/2}$, $\Lambda_1= \sqrt{2f_Q}\lambda+\frac{f_R}{2}$, 
 $\Lambda_2= \sqrt{2f_Q}\sigma$, $\lambda= \frac{1}{8}\left[3\sqrt{2f_Q}\left(R+\frac{f_R}{f_Q}\right)\pm\sqrt{{2f_Q}\left(R+\frac{f_R}{f_Q}\right)^2-8\kappa^2(\rho+P)}\right]$, and $\sigma=\lambda\pm\sqrt{\lambda^2-\kappa^2(\rho+P)}$.
In terms of the auxiliary metric $h_{\mu\nu}$, the field equations (\ref{eq:met-varX}) for the metric can be written as
\begin{equation}\label{eq:Tmn-pfh}
R_{\mu\nu}(h)=\frac{1}{\Lambda_1}\left[\frac{\left(f+2\kappa^2\Pi\right)}{2\Omega}h_{\mu\nu}+\frac{\Lambda_1\kappa^2(\rho+\Pi)}{\Lambda_1-\Lambda_2}u_{\mu}u_\nu\right]
\end{equation}
In order to study physical predictions of these theories, we must consider specific models for which the functions  $R=\R(\rho,\Pi)$ and $Q=\Q(\rho,\Pi)$ can be solved. An example is provided by the family of Lagrangians $f(R,Q)=\tilde{f}(R)+b{Q}/{R_P}$. 
In this case, we find that $\R=\R(T)$ and $\Q$ is given by     
\begin{equation}\label{eq:lambda2}
\frac{\Q}{2R_P}=-\left(\kappa^2\Pi+\frac{\tilde f}{2}+\frac{R_P}{8}\tilde f_R^2\right)+\frac{{R_P}}{32}\left[3\left(\frac{R}{R_P}+\tilde f_R\right) \pm\sqrt{\left(\frac{R}{R_P}+\tilde f_R\right)^2-\frac{4\kappa^2(\rho+\Pi)}{R_P}}\right]^2
\end{equation}
For the sake of clarity, let us consider the quadratic Lagrangians $f(R,Q)=R+aR^2/R_P+Q/R_P$, for which $R$ turns out to go exactly like in GR, $\R=-\kappa^2T$ (the stellar structure of quadratic $f(R)$ models has been discussed recently\cite{quadratic}). If in this example we choose $a=-1/2$, i.e., $\tilde{f}(R)=R-R^2/(2R_P)$, then (\ref{eq:lambda2}) becomes
\begin{equation}\label{eq:Q-1/2}
\Q=\frac{3R_P^2}{8}\left[1-\frac{2\kappa^2(\rho+\Pi)}{R_P}+\frac{2\kappa^4(\rho-3\Pi)^2}{3R_P^2}-\sqrt{1-\frac{4\kappa^2(\rho+\Pi)}{R_P}}\right] \ .
\end{equation}
At low energies, this expression recovers the GR limit 
\begin{equation}
Q\approx \left(3 P^2+\rho ^2\right)+\frac{3 (P+\rho )^3}{2 R_P}+\frac{15 (P+\rho )^4}{4 {R_P}^2}+\ldots
\end{equation}
However, positivity of the argument in the square root of (\ref{eq:Q-1/2}) implies that 
$\kappa^2(\rho+\Pi)\leq {R_P}/{4}$, 
which clearly shows that the combination $\rho+\Pi$ is bounded from above. \\
\indent We have shown that the independent connection can be expressed as the Levi-Civita connection of an auxiliary metric which is related to the physical metric $g_{\mu\nu}$ and the energy-momentum tensor by means of a non-standard transformation, which becomes disformal when matter is described as a perfect fluid and boils down to conformal when the $R_{\mu\nu}R^{\mu\nu}$ dependence disappears. The emergence of two metrics related by a disformal transformation, a basic requirement of relativistic MOND  \cite{Bekenstein04} theories to properly account for gravitational lensing, could make these theories interesting for the consideration of dark-matter-related problems. \\
\indent We have also shown that in Palatini $f(R,R_{\mu\nu}R^{\mu\nu})$ theories the scalars $R$ and $R_{\mu\nu}R^{\mu\nu}$ can in general be written as functions of $\rho$ and $\Pi$ but not necessarily via the trace $T$. As a result, the phenomenology of these theories is much richer than that of the individual $f(R)$ or $f(R_{\mu\nu}R^{\mu\nu})$ theories. Furthermore, for some simple models, we have shown explicitly that the scalar $R_{\mu\nu}R^{\mu\nu}$ sets bounds on the physically accessible range of $\rho$ and $\Pi$, which suggests that scenarios such as the very early Universe and the last stages of stellar collapse could be seriously affected by the new dynamics possibly leading to singularity resolution\cite{BOSA09,Olmo09}. \\
\indent The results obtained in this work open new avenues of research in the context of the very early Universe, the radiation dominated epoch, and the accelerating Universe, with new mechanisms to generate an effective cosmological constant, within a framework that does not introduce new degrees of freedom and, therefore, is closer to GR than other type of modified theories of gravity or dark energy models. \\
\indent {Acknowledgments.} H.S-A. has been partially supported by the Austrian Science
Fund FWF under Project No. P20592-N16. G.J.O. thanks MICINN for a Juan de la Cierva contract, the Spanish Ministry of Education and its program ``José Castillejo'' for funding a stay at the CGC of the University of Wisconsin-Milwaukee, and the Physics departments of the UW-Milwaukee and UW-Barron County for their hospitality during the elaboration of this work. G.J.O. has also been partially supported by grant FIS2008-06078-C03-02.

\end{document}